# Do We Observe Dark Photons in PHELEX with a Multi-Cathode Counter ?

**Anatoly Kopylov\*, Igor Orekhov and Valery Petukhov**

Institute for Nuclear Research, Russian Academy of Sciences, 117312 Moscow, Russia
\* Correspondence: kopylov@inr.ru

**Abstract:** In this paper, we report the results obtained in the PHoton–ELectron EXperiment (PHELEX) during 208 days of measurements in the search for dark photons using a multi-cathode counter with an iron cathode. We observed a systematic excess of single-electron events in four runs, with a confidence level better than $5\sigma$, in the time interval from 8-00 to 12-00 in sidereal time. The fact that we observe this effect in the stellar frame while detecting the excess of count rate in the solar frame in different time intervals is evidence that this effect is of galactic origin. This strengthens our argument that we are indeed observing the effect of dark photons.

## 1. Introduction

Using a specially designed by us multi-cathode counter, we have developed a technique to search for dark photons [1]. This technique is based on detecting single electrons emitted from a metallic cathode during the conversion of dark photons, which can be considered as particle-like solutions with the flux of magnetic or electric fields [2, 3], into single electrons emitted from the cathode's surface. This method efficiently sets upper limits for the kinetic mixing parameter [4]. An important feature of this method is that it utilizes free electrons of a degenerate electron gas in a metal cathode as a target. In a previous study [5], it was discussed that if dark photons are polarized (i.e., if the vector of the **B** or **E** field of dark photons has a certain direction in the stellar or solar frame), one can observe diurnal variations in the count rate of single electrons in the corresponding frame owing to the rotation of the Earth. In this study, we adopted a reference model for the angular dependence of the effect from dark photons, which assumes a $\cos^2\theta$ relationship, where $\theta$ represents the angle between the vector of the **B** (**E**) field and the cathode's surface. However, the real angular dependence may differ because we currently lack a robust theory of dark photons, including knowledge of their exact field configuration [2, 3]. This needs to be verified experimentally. Regardless of the actual angular dependence, the curve of diurnal variations should be symmetrical with respect to some moment in time when the vector of the **B** or **E** field of dark photons lies in the plane of a meridian where the detector is located. This symmetry arises from the effect of the Earth's rotation. Dark photons, possessing mass, are presumed to have both transverse and longitudinal modes for the **B** (**E**) field. If the motion of dark photons is ordered due to possible interactions among dark photons themselves, this may result in a certain polarization, at least for the longitudinal mode. We will study this phenomenon in the future. The observation that the effect of some excess in count rate at certain time interval is present in the stellar frame but absent in the solar frame provides evidence of its galactic origin, strengthening the argument that dark photons likely cause the observed effect. In this paper, we present the results of our experiment, which seeks to detect diurnal variations in the count rate of single electrons using a multi-cathode counter with an iron cathode.

## 2. Materials and methods

The objective of this experiment was to observe potential diurnal variations in the count rate of single electrons resulting from the conversion of dark photons on the cathode's surface [6]. We used a gaseous proportional counter with a cylindrical iron cathode 166 mm in diameter and 500 mm in length. The cathode was encapsulated in a stainless-steel housing, with a quartz window for calibration using a UV lamp. The counter was filled with a mixture of $Ne + CH_4$ (10%) at 0.1 MPa. The detector was placed on the ground floor of a building in Troitsk, Moscow, Russia, in a specialized cabinet equipped with a 30-cm steel shield and 10 cm of boronated polyethylene. The steel shield protected against external gamma radiation, while the boronated polyethylene shielded against thermal neutrons. The counter was placed horizontal-ly, with the cathode's axes positioned at a 23° angle relative to the North–South direction [7]. A 25-μm-thick gold-plated tungsten–rhenium wire was stretched along the cathode's axis, serving as the anode for the counter. Three cylindrical cathodes were arranged around this wire [1]. The first cathode, with a 40-mm diameter, encircled the anode wire. The second cathode was positioned immediately adjacent to the third cathode (the outer iron cathode) and located 8 mm from the third cathode's surface. The first and second cathodes were constructed from 50-μm-thick gold-plated tungsten–rhenium wires, tighten along the cathode with an approximately 5-mm pitch. For a more detailed description of the counter, refer to a previously published article [1]. All three cathodes were maintained at high voltage. The first cathode, with a 40-mm diameter, provided high gas amplification. The second one, positioned 8 mm from the surface of the outer iron cathode, served as a barrier for electrons emitted from the iron cathode. In the first configuration, when the potential at the second cathode was higher than that of the outer cathode electrons passed through this barrier to reach the anode. In the second configuration, when the potential of the second cathode was lower than that of the outer cathode, electrons were repelled back to the surface of the cathode. In the first configuration, we counted electrons emitted from the iron cathode's surface and electrons generated in the gas within the counter's volume. In the second configuration, only electrons generated in the gas were counted. The signal from the counter's anode was sent to the input of a charge-sensitive preamplifier, and the preamplifier's output voltage was then processed by an ADC board [1]. The counter was calibrated using UV radiation from a mercury vapor lamp. The signal from the preamplifier's output was digitized by an ADC board with a sampling rate of 10 MHz, resulting in 1 TB of data per day. The operational principle is illustrated in Figure 1, which is taken from a previously published article [5]. Small single-electron pulses are often positioned on the baseline below the zero line. Continuous recording was necessary to recover the pulse's shape atop this varying baseline. This explains why we accumulated such substantial amounts of data daily. We collected pulses with amplitudes ranging from 3 to 50 mV, characterized by a "good" shape, indicative of "true" single-electron pulses. The electronic noise level remained at 400 μV.

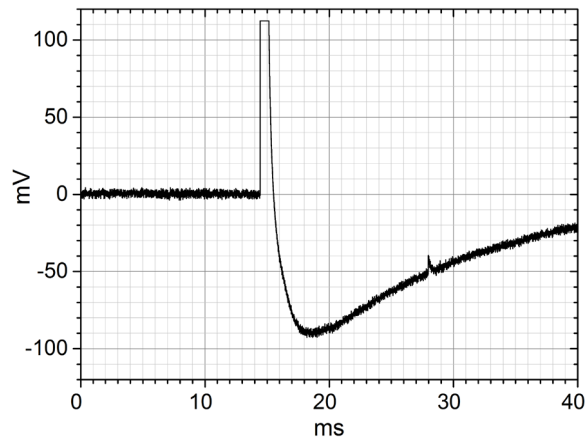

Figure 1. One of the snapshots recorded during the measurements. A small single-electron pulse trailing the larger pulse generated by a muon passing through the counter is observed.

Data analysis was conducted offline, with a detailed description provided in previously published articles [1, 4]. The recorded count rates were categorized into 2-h intervals in both sidereal and terrestrial times. If the vector of the **B** or **E** field assumed a specific orientation in the stellar or solar frame, diurnal variations could be observed in the corresponding frame owing to the Earth's rotation. This would be a compelling argument supporting the hypothesis that the observed effect is attributed to dark photons. Figure 2 shows the results obtained from four measurement runs in sidereal time, while Figure 3 illustrates the same results in terrestrial time. Each run spanned 52 days and comprised 544 data points. Within each data point, 270 ÷ 370 single-electron events were collected across 5000 frames, each lasting 0.2 s. Thus, each data point in Figures 2 and 3 encompasses 12000 ÷ 16000 single-electron events. Each frame contained two million records for each 100-ns timeslot.

## 3. Results

Figure 2 illustrates that we observed a systematic excess of events in all four runs during the time interval from 8-00 to 12-00 in sidereal time.

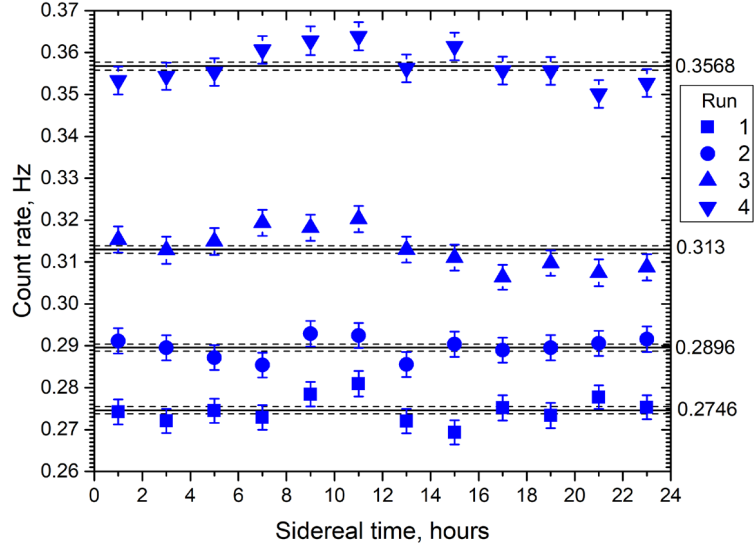

Figure 2. Diurnal variations obtained from four runs in sidereal time. Solid lines represent the average values, and dashed lines indicate ±σ for the average values.

The fact that we consistently observed this excess in all four runs during the same time interval, from 8-00 to 12-00, makes it highly improbable that this result is simply a product of pure statistical fluctuation. In all four runs, this event—where we observed an excess of more than (or equal to) one sigma above the average value in two adjacent sidereal time intervals—is not visible anywhere else except within the 8-00 to 12-00 sidereal time interval. Obviously, the possibility of obtaining a similar excess in all four runs during the same time interval appears unlikely, if to do Monte Carlo simulations. The significance of this effect was assessed by calculating the probability using the following expression:

$$p = 12 \prod_{i=1}^{2n} \left( 0,5 \, erfc\left( \frac{x_i}{\sqrt{2}} \right) \right) \qquad (1)$$

Here, $p$ represents the probability of obtaining the observed temporal pattern of the excess in two adjacent bins across $n$ runs, with $x_i$ denoting the excess value of the count rate in the $i$-th bin in units of σ over the average value. In each run, we counted rates in only two bins: the first for the interval in sidereal time from 8-00 to 10-00 and the second for the interval from 10-00 to 12-00. The calculated probability that this result is a product of pure statistical fluctuation has been determined to be $2 \times 10^{-10}$, corresponding to a confidence level better than 5σ.

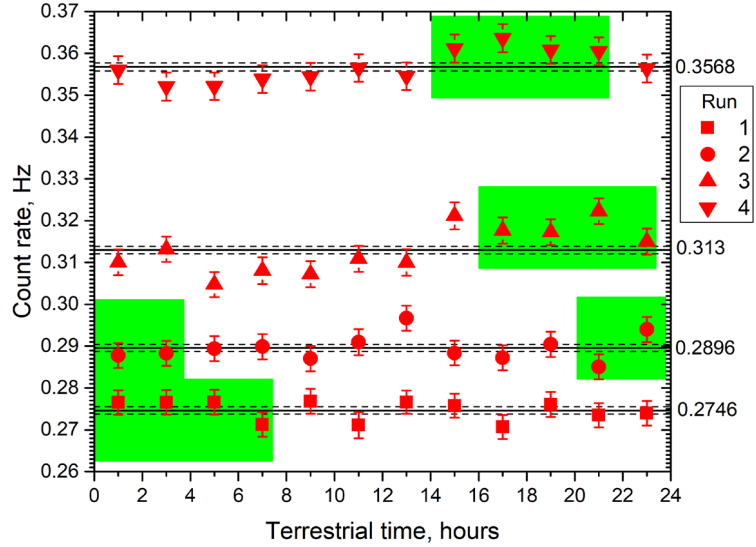

Figure 3. Diurnal variations obtained in four runs in terrestrial time. Green zones represent the events obtained in the actual runs contributing to the time interval from 8-00 to 12-00 sidereal time.

We do not observe a similar feature in Figure 3 in terrestrial time. This provides strong evidence that the effect observed in Figure 2 during the time interval from 8-00 to 12-00 is highly likely to be caused by dark photons. In time intervals in terrestrial time corresponding to the interval from 8-00 to 12-00 sidereal time, we also observed an excess in the count rate. But here there's no coincidence in time and the duration of each terrestrial time interval is 7 h 20 min (4 h of sidereal interval and 3 h 20 min of the time shifting during 52 days of measurements). As one can see from Fig. 2 and Fig. 3 the effect seems to be more pronounced in runs 3 and 4 than in runs 1 and 2. The first two runs has been conducted in the first quarter of the year, the second two runs – in the second quarter of the year. It will be interesting to see if these results are reproduced. The increase in the average count rate from runs 1 to 4 can be attributed to the rise in ambient temperature. This increase results from the higher emission rate of thermal electrons, presumably originating from impurities in the cathode metal. The last run occurred in July, whereas the first occurred in January. By comparing the average rates across different runs, the effect of the sharp temperature increase in the Moscow region in July becomes evident.

We previously performed measurements in two configurations [1, 4]: the first configuration involved the detection of electrons emitted from the cathode's surface and electrons generated in the gas within the counter's volume. In contrast, the second configuration only detected electrons generated in the gas within the counter's volume. We determined the upper limits for the constant of kinetic mixing by comparing the count rates obtained in these two configurations. However, when observing diurnal variations, there is no need to compare count rates between two configurations. In this case, we focus solely on the first configuration and search for an excess in the count rate during specific time intervals. Naturally, a pertinent question arises: how can we be certain that the source of this excess is electrons emitted from the cathode's surface rather than electrons generated in the gas within the counter's volume? We exclude the latter possibility because we cannot propose any plausible mechanism of galactic origin for diurnal variations in the latter case. Nevertheless, even as we observe an effect presumably caused by dark photons, confirming whether it originates from the cathode's surface or the gas within the counter's volume remains essential. Recently, we have obtained results from measurements conducted in the second configuration, albeit with limited statistics from a single run lasting 52 days. As shown in Figure 4, the possibility that the source of the effect is electrons generated in the counter's volume is highly unlikely.

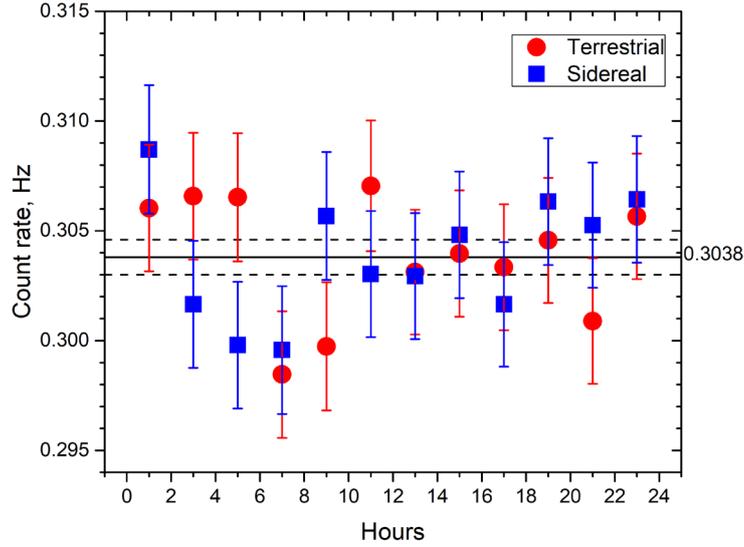

Figure 4. Diurnal variations observed during measurements conducted over 52 days in the second configuration, presented in sidereal time (blue) and terrestrial time (red)

At the very least, we did not observe any excess during the time interval from 8-00 to 12-00 sidereal time in this case. However, to state this with complete certainty, we require a similar dataset involving approximately 3 ÷ 4 runs of measurements, as we had in the first configuration. We are currently continuing measurements in the second configuration, and we anticipate definitively resolving this question by the end of this year.

We still have an open question regarding whether the effect depends on the type of metal used for the counter's cathode. In this study, we utilized an iron cathode, a magnetic material. In future experiments, we plan to employ a cathode made of aluminum, a non-magnetic material. It will be interesting to investigate whether the effect observed in the stellar frame, as presented in Figure 2 with the counter featuring an iron cathode, will also be observed when using an aluminum cathode.

What steps must be taken to verify this result? We have observed the effect as an excess count rate of $\Delta R = (0.006 \pm 0.001)$ Hz during the time interval from 8-00 to 12-00, above the background caused by thermal electrons and electrons resulting from natural radioactivity when averaged over the remainder of the sidereal time. What can be said about the parameter of kinetic mixing? The upper limit for this parameter obtained in our previous measurements [8] was about $3 \cdot 10^{-12}$ for a mass of dark photon in the range from 9 to 40 eV. This upper limit for parameter of kinetic mixing was obtained from the upper limit of count rate 0.003 Hz. If to average the value $\Delta R = (0.006 \pm 0.001)$ Hz for all day we obtain 0.001 Hz. From here we can obtain, as a crude estimate, the value for the parameter of kinetic mixing in our case on the level of $10^{-12}$. Of course, this is just the "order of magnitude" estimate. The upper limits obtained in other experiments (e. g. reference [9] and others therein) for this mass range is lower, but here we again must emphasize that in PHELEX we use as a target free electrons of a degenerate electron gas of a metal. This may account for the reason why we observe the effect on this level of kinetic mixing. For further progress it is vital to demonstrate that this result is reproducible. We need to conduct four more runs in the first configuration to achieve this.

Why we observe one peak, not two? Figure 5 explains this. If the vector of polarization during maximum of the peak is perpendicular to the axis of the counter, the effect is maximal. For our geographical latitude it means that this vector is close to $45^0$ to the axis of the Earth. In 12 hours, this vector will be approximately along the axes of the counter and the effect is minimal. So, the fact that we observe only one peak means that the vector of polarization is close to $45^0$ to the axes of the Earth when this vector is in the plane of the Moscow meridian.

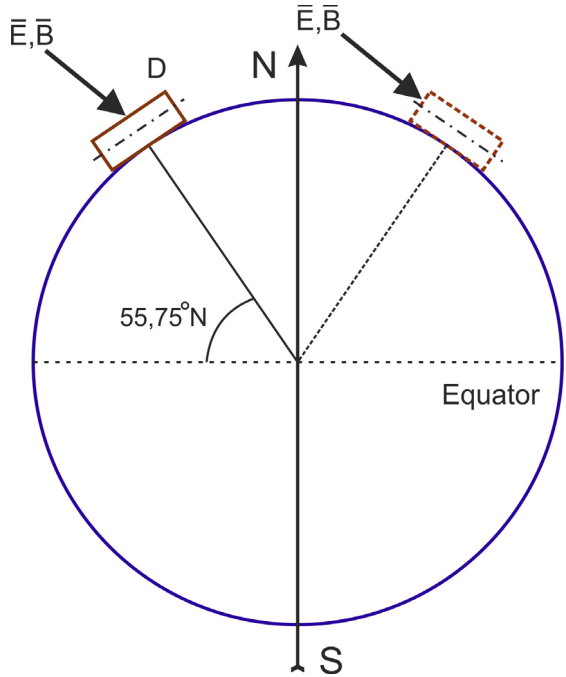

Figure 5. The possible direction of **E**, **B** fields relative to the axes of the counter D.

To establish definitively that this effect is indeed attributable to dark photons, similar measurements should be performed in a laboratory with the same geographical latitude but different geographical longitudes. For example, if we were to conduct this experiment near Manchester, which shares a similar geographical latitude but has a time difference of two hours relative to Moscow, we would expect to observe a similar effect during the time interval from 10-00 to 14-00 in sidereal time. Thus, the most efficient approach to expedite the verification process is to perform similar measurements in diverse laboratories situated at different locations and employ counters with cathodes made of different metals. The effect should vary depending on the counter's orientation (i.e., vertical, horizontal, with axes oriented East–West or North–South) and the geographical latitude and longitude of the site where the detector is located. Notably, this effect can only be observed when the cathode of the counter is polished to a mirror finish.

### 4. Conclusions

We collected data from four separate runs, each over approximately 52 days, and consistently observed a systematic excess of count rate occurring between 8-00 and 12-00 sidereal time during each run. The fact that we observe this effect in the stellar frame while detecting the excess of count rate in the solar frame in different time intervals is evidence that this effect is of galactic origin. This strengthens our argument that we are indeed observing the effect of dark photons. The probability that this result is simply a product of pure statistical fluctuation across all four runs is estimated to be $2 \times 10^{-10}$ that correspond to confidence level better than $5\sigma$. We are continuing to collect additional data to understand the details and achieve a more accurate interpretation of the results.

**Acknowledgments:** We appreciate very much the substantial support from the Ministry of Science and Higher Education of Russian Federation within "The Instrument Base Renewal Program" in the framework of the State program "Science".